\documentclass[english]{article}
\usepackage[T1]{fontenc}
\usepackage{float}
\usepackage{graphicx}
\usepackage{esint}

\makeatletter

\providecommand{\tabularnewline}{\\}

\newcommand{\lyxaddress}[1]{
	\par {\raggedright #1
	\vspace{1.4em}
	\noindent\par}
}

\usepackage{babel}
\usepackage[superscript,biblabel]{cite}

\makeatother

\usepackage{babel}
\begin{document}
\title{Site-site interaction model for alcohol models in two-dimensions}
\author{Aur\'{e}lien Perera\thanks{email: aurelien.perera@sorbonne-universite.fr}}
\maketitle

\lyxaddress{Laboratoire de Physique Th\'{e}orique de la Mati\'{e}re Condens\'{}e (UMR CNRSe
7600), Sorbonne Universit\'{e}, 4 Place Jussieu, F75252, Paris cedex 05,
France.}
\begin{abstract}
An interaction site-based model of two-dimensional alcohols is proposed
as a follow up of the recent SSMB site-site model for 2D water {[}\emph{J.
Mol. Liq. 386 (2023 122475}{]}. Computer simulation studies indicate
that the model exhibits hbond-type clustering based on the same charge
order feature observed in real alcohols. Hence, the equivalent of
2D mono-ols ranging from methanol to octanol were studied for their
clustering properties, focusing on how the micro-structure affects
the shape of the site-site pair correlation functions and structure
factors, as well as the combination of the latter into the radiation
scattering intensities. The major finding is the apparent contradiction
between the existence of large pre-peaks in the structure factors,
usually associated to the existence of clusters, and the exponential
decay of the cluster distribution indicating the absence of specific
clusters, contrary to the 3D case. This is resolved by realizing that
the pair correlation function is an observable of the local density
fluctuations, hence the pre-peak witnesses fluctuations around clustering
tendencies, which are the result of charge ordering of the polar groups,
and visible in the snapshots. The scattering pre-peak witnesses only
fluctuations due to charge ordering, and not the clusters themselves,
underlining the fact that these are labile entities. The study highlights
how charge order through atomic sites is a universal feature behind
the micro-structure of organized liquids, and, in the particular case
of 2D liquids, a more realistic alternative to orientation based models
such as the Mercedes-Benz model, for instance.
\end{abstract}

\section{Introduction\label{sec:Introduction}}

Since the introduction of two-dimensional Ising model \cite{Ising1925},
our understanding of the relationship between the microscopic properties
of matter and the macroscopic properties have have been greatly improved,
particularly in what concerns phase transitions\cite{landau2013statistical,chaikin2000principles}.
For realistic liquids, two-dimensional models of hard discs \cite{Alder1962}
and Lennard-Jones liquids\cite{2D-LJ} have merely served as mimics
of their three-dimension counter-parts, serving mostly to test simulations
and theories. However, liquids can be classified into two categories;
simple and complex liquids \cite{myIUPAC}. Liquid nitrogen is a canonical
example of simple liquid that can be modeled by a Lennard-Jones atom\cite{HansMac}.
These liquids have greatly helped improve our theoretical approach
based on statistical mechanics, which was once considered as impossible
by Lev Landau himself, as cited by P. G. de Gennes in \cite{deGennes_on_Landau}.
But this is not the end of the story, when complex associating liquids
such as water come into play. Indeed, water is an hydrogen bonding
liquid\cite{Hbond} and this directional bonding gives water the status
of special liquid, with more than 60 anomalous properties in the liquid
phase alone. Because of this special interaction, water remains a
liquid which is considered as not fully understood from microscopic
point of view, both as a single component and more so when it is in
mixing conditions\cite{Finney2024}. In particular, our theoretical
understanding from statistical physics point of view is still in infancy.
A quick view of why this is so, can be obtained by modeling the hydrogen
bonding property by a simple classical Coulomb charge association.
This is possible by attributing a negative charge to the oxygen atom
(usually around -0.7e) and a positive charge to the each of the two
hydrogen atoms (usually around -0.3e). Coulomb charge association
being about 557 times greater than the van de Waals interaction at
distances above molecular contact, in dense liquid phase context,
one has labile cluster formation which give water a two state dynamics,
that of free and bound molecules. This two state model was formulated
on a macroscopic basis by R\"{o}ntgen and corresponds to one of the earlier
models of water. In this context, unlike simple liquids, it is really
useful to have a two-dimensional counterpart, in order to disentangle
the various contributions between microscopic and macroscopic properties,
which make water a liquid difficult to fully understand. The Mercedes-Benz
(MB) model\cite{BenNaim-MB,DillReview} have been a good example of
such two-dimension model.

The MB model is purely orientational, similar to the dipole representation
of water. We have recently introduced a 2D water model based on charged
site representation, the SSMB model\cite{SS-water} , which mimics
real water on the basis of atom-atom interaction. While very similar
to the original MB model, it was found to produce a more realistic
pair correlation function, similar to that of real water, and more
importantly, a pair structure factor with the characteristic shoulder
peak of the x-ray spectra of real water\cite{AmannWinkel2016,water-3PS},
which is absent for the MB model. This is an important point because
it witness the microscopic dual state of water, not in real space,
as postulated by the MB model, but in the reciprocal space, as found
for the real water. In addition, it should be noted that the pair
correlation function plays a capital role in the statistical physics
based description of disordered liquids, unlike ordered liquids (such
as in the Ising model for instance), where it is the one body correlation
function which correspond to the order parameter. This function describes
better the local disorder, and is a capital indicator of the difference
between simple and complex liquids.

Alcohols have played an important role in the comprehension of complex
liquids, since these liquids tend to form local hydrogen bonded clusters
which are visible from the x-ray scattering spectra, through the characteristic
pre-peak in the range $k\approx0.5-1\mathring{A}^{-1}$ \cite{my_monool},
unlike water, for which such clusters exists, but do not produce a
well separated pre-peak, but merely a shoulder peak\cite{water-3PS,water-xray-Hura}.
This is very intriguing, since hydrogen bonding in water plays a more
important role than it does for alcohols, even though both are associating
liquids. For instance, alcohols are not know to possess the many anomalous
properties that water does. In this context, it appears as necessary
to have a site interaction 2D model of alcohols, as a complement to
that of water. Indeed, an equivalent of the MB model for alcohols
have been introduced previously \cite{alc-old} and a structural analysis
of the model was presented recently by Urbic \cite{alcTomaz}. 

The present work aims at extending the MB-like model of alcohols to
a site-site representation, which would be an equivalent of the SSMB
model to the MB water model. Since our calculations are in the Canonical
Ensemble at constant NVT, as opposed to those of Ref.\cite{alcTomaz}
in the isobaric ensemble with constant NPT, we impose the same packing
fraction in order to compare different alcohols. In addition, we compute
the site-site structure factors and the equivalent of the scattering
intensity, which is one way to validate the models by comparing the
pre-peaks and their dependence in alkyl tail length and temperature
to that obtained from experiments\cite{my_monool} .

\section{Model, theoretical and computational details\label{sec:Model,-theoretical-and}}

\subsection{2D alcohol models\label{subsec:2D-alcohol-models}}

\begin{figure}[H]
\includegraphics[scale=0.3]{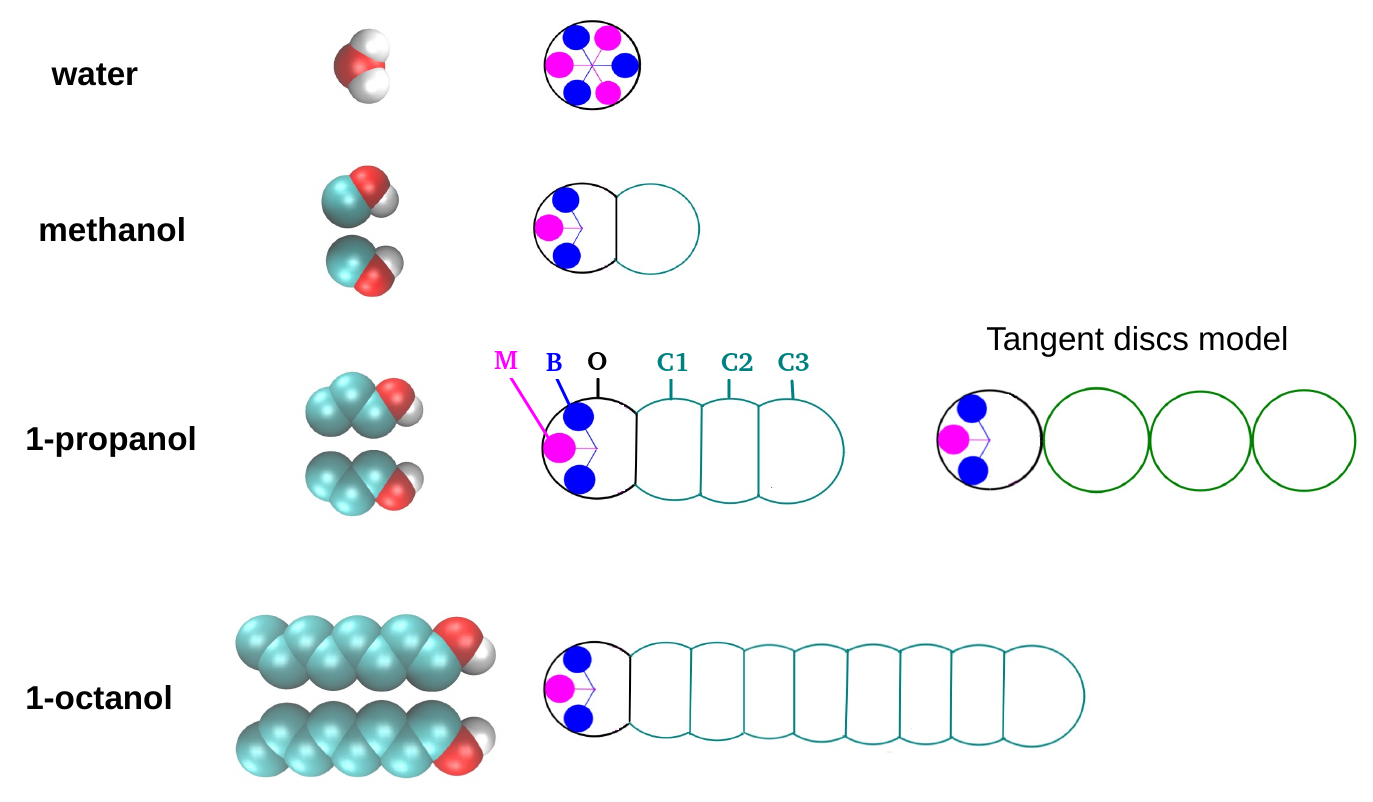}

\caption{Illustration of the 2D non-chiral alcohol models, compared to their
3D/2D realistic chiral models (see text). The site-site equivalent
of 2D 1-propanol model of tangent discs of Ref.\cite{alcTomaz} is
shown for comparison. The site-site water model of Ref.\cite{SS-water}
is equally shown for comparison. For the real molecules, oxygen site
is red, hydrogen in white and methyl/methylene site in green. For
the 2D acohols, the hydroxyl head is black circle, the attractive
sites are in blue, repulsive in magenta, and methyl sites are green
circles. The sites positions are explained in the text. The nomenclature
for naming site is illustrated for the propanol molecule.}
\label{FigModels}
\end{figure}

The interaction site based two dimensional alcohol models follow the
same patterns we have previously developed in the case of 2D site-site
SSMB model of water \cite{SS-water}. Namely, we replace the interaction
between the MB arms by ``charged'' site interactions, combined with
a repulsive interaction site. This is illustrated in Fig.\ref{FigModels}.

The hydroxyl head group is modeled as a 2 arms state, with 2 attractive
sites (in blue in Fig.\ref{FigModels}), which represent both the
attractive oxygen and hydrogen sites. We have followed Ref.\cite{SS-water}
and made the two attractive site of equal valence. This is also consistent
with the MB symmetry of 2D water model, for which the oxygen and hydrogen
sites are not distinguished and simply replaced by similar attractive
sites, and the 3 arms should be identical. This seemingly unphysical
equivalent of the real water is justified by the fact that such simple
model is able to capture the physical properties of the associating
liquids as well as some of the anomalies of real water, such as the
maximum density, the minimum of compressibility and the two types
of amorphous phases (dense HDL and expanded LDL)\cite{SS-water}.
In order to avoid misalignments of the 2 arms, it is necessary to
introduce a single repulsive site (in magenta in Fig.\ref{FigModels}).
The interactions have been taken similarly to our previous work for
water, with a $1/r^{12}$ repulsion and a Yukawa form for hydrogen
bonding part of the interaction between two atomic sites $a$ and
$b$:
\begin{equation}
v_{ab}(r)=4\epsilon_{ab}\left(\frac{\sigma_{ab}}{r}\right)^{12}+557\alpha_{ab}\frac{Z_{a}Z_{b}}{r}\exp\left[-\frac{(r-R_{ab})^{2}}{\kappa_{ab}}\right]\label{V12}
\end{equation}
where we use the Lorentz rule for the site diameters $\sigma_{ab}=(\sigma_{a}+\sigma_{b})/2$
and the Berthelot rule for the energy parameters $\epsilon_{ab}=\sqrt{\epsilon_{a}\epsilon_{b}}$,
where $\sigma_{a}$ and $\epsilon_{a}$ are the diameter and energy
of site $a$, respectively. In what concerns the screened Coulomb
Yukawa part $Z_{a}$ is the valence of site $a$, and $\alpha_{ab}$
is the sign of the interaction between sites $a$ and $b$. As can
see, in Eq.(\ref{V12}), the second term is not really Coulomb like
since $\alpha_{ab}\neq\alpha_{a}\alpha_{b}$. These coefficients are
listed in the last row of Table 1. The coefficient 557 results from
reducing the magnitude of the 3D Coulomb interaction, as was explained
in Ref.\cite{CO1}. It is because of the large discrepancy of magnitude
between the first and second term, that charged and uncharged atomic
groups behave differently and produce the typical micro-structure
seen in associated liquids.

The alkyl tail atoms are modeled as neutral zero valence sites $Z_{C}=0$.
The position of these ``carbon'' atoms is decided using the following
criteria. If one is to respect the shifted positions as in real alcohol
models, as in the left column of Fig.\ref{FigModels}), then the 2D
version model become chiral. This is illustrated by showing the two
possible symmetrical configurations, which become distinct in 2D.
To avoid this unnecessary features, we have modeled the alkyl tail
as a linear sequence of carbon atoms. However, contrary to Ref.\cite{alcTomaz},
where the discs are tangent, we have chosen to merge them up to half
diameter, except for the first carbon, which is placed at a higher
distance, such that the hydroxyl head is more exposed, avoiding hindrance
of Hbond. All outer atoms have been given the same diameter $\sigma_{i>3}=\sigma_{1}=\sigma$
, that of the hydroxyl head. Inner atoms of the hydroxyl head (blue
and magenta sites) have the same diameter $\sigma_{B}=\sigma_{\mbox{blue}}=\sigma_{2}=\sigma_{M}=\sigma_{\mbox{magenta}}=\sigma_{3}=\sigma/3$.

The parameter of the alcohol models are summarized in the following
Table 1 and Table 2.

\begin{table}[H]
\begin{centering}
\begin{tabular}{|c|c|c|c|c|c|c|c|c|c|c|c|c|}
\hline 
$\sigma$ & $\sigma_{B}$ & $\sigma_{M}$ & $\sigma_{C}$ & $\epsilon$ & $\epsilon_{B}$ & $\epsilon_{M}$ & $\epsilon_{C}$ & $Z_{B}$ & $Z_{M}$ & $Z_{C}$ & $\kappa$ & $R_{i}$\tabularnewline
\hline 
\hline 
$1$ & $1/3$ & $1/3$ & $1$ & $1$ & $1/2$ & $1/2$ & $1$ & $0.3$ & $0.15$ & 0 & $0.15$ & $0$\tabularnewline
\hline 
\multicolumn{10}{|c|}{$\alpha_{PP}=-1$;$\alpha_{PM}=1$;$\alpha_{MM}=1$; all other $\alpha_{ab}=0$} & \multicolumn{3}{c|}{}\tabularnewline
\hline 
\end{tabular}
\par\end{centering}
\caption{Interaction parameters for 2D alcohols as in Eq.(\ref{V12})}
\end{table}

\begin{table}[H]
\begin{centering}
\begin{tabular}{|c|c|c|c|c|c|}
\hline 
 & $O$ & $B$ & M & $C_{1}$ & $C_{n}$ ($n=2-8$)\tabularnewline
\hline 
\hline 
$r$ & 0 & $1/3$ & $1/3$ & $3/4$ & $3/4+(n-1)n$\tabularnewline
\hline 
$\theta$ & 0 & $\pm\pi/3$ & $0$ & $\pi$ & $\pi$\tabularnewline
\hline 
\end{tabular}
\par\end{centering}
\caption{Site polar coordinates for 2D alcohols}
\end{table}

\subsection{Computational and theoretical details\label{subsec:Computational-and-theoretical}}

Standard Monte Carlo simulation have been used in the constant N,
V, T Canonical ensemble, a choice justified by the necessity to study
structural properties with fixed number density $\rho_{N}=N/V$. The
reduced density and temperature are defined in Lennard-Jones units
related to the parameters of the polar head site O as $\rho=(N/V)\sigma^{2}$
and $T=k_{B}T_{K}/\epsilon$ (where $T_{K}$ is the temperature in
Kelvin). Three temperatures are studied, $T=1$, $T=2$ and $T=3$,
which would correspond to liquid phases above the triple point. The
densities vary between $\rho=0.1$ and $\rho=0.45$, the latter accessible
only for methanol. A more appropriate density would the packing fraction
$\eta=\rho S_{n}$, where $S_{n}$ is the surface of the molecule
which depends on the number of monomers $n$ ($n=1-8$)

\begin{equation}
S_{n}=(n+1)\frac{\pi}{4}\sigma^{2}-2L\left(\frac{\sigma}{8}\right)-2nL\left(\frac{\sigma}{4}\right)\label{Surf}
\end{equation}
where $L(d)$ is the surface of the lens of a disc of diameter $\sigma$
at distance $d$ from the center.
\begin{equation}
L(d)=\frac{\sigma}{4}^{2}\arccos\left(\frac{2d}{\sigma}\right)-\frac{d\sigma}{2}\sqrt{1-\left(\frac{2d}{\sigma}\right)^{2}}\label{lend}
\end{equation}
An alternative packing fraction $\eta^{(DR)}=\rho S_{n}^{(DR)}$ would
be to scale according to the surface of the disco-rectangle $S_{n}^{(DR)}$
which contains the molecule
\begin{equation}
S_{n}^{(DR)}=\frac{\sigma^{2}}{4}(\pi+1+2n)\label{S_DR}
\end{equation}
The system sizes are taken to be large enough to ensure all site-site
correlation to decay to the asymptote $g_{ab}(r)\rightarrow1$ by
$r=(1/3)L$ where $L$ is the box size. This corresponds to $N=700$
for methanol to $N=336$ for octanol. The systems are started with
square box configurations were all the particles are aligned. The
number of particle is determined by the condition for a rectangle
containing at most one particle to pave the entire box. Equilibration
and production are about $50\times10^{3}$ steps, where one step consists
in attempting to move all N particles successively. In the case of
low temperature and longer alcohols, it is necessary to use 100 thousand
steps. All pairs of site-site correlation functions $g_{ab}(r)$ between
pairs of sites ($a,b$), as well as internal energies and virial pressures
are evaluated during the production runs. The site-site structure
factors are obtained by a 2D Fourier transform of the pair correlation
functions:
\begin{equation}
S_{ab}(k)=1+\rho\int d\mathbf{r}\left[g_{ab}(r)-1\right]\exp(i\mathbf{k.r})\label{S(k)}
\end{equation}
The total structure factor is equally required
\begin{equation}
S_{ab}^{(T)}(k)=w_{ab}(k)+S_{ab}(k)-1\label{ST(k)}
\end{equation}
which involves the intra-molecular correlations, which for rigid molecules
are defined as
\begin{equation}
w_{ab}(k)=j_{0}(kd_{ab})\label{intra}
\end{equation}
where $d_{ab}$ is the distance between sites $a$ and $b$ inside
the molecule. The 2D Fourier transforms are numerically evaluated
by Talman transform techniques \cite{TALMAN,my_2D_elli}. This step
necessitate to convert the correlation functions $g_{ab}(r)$ from
equal r-steps to log-scale. This is dont using a linear interpolation,
which adds small oscillatory artifacts visible in the k-range below
$k\sigma\approx0.05$. Finally, the scattering intensity is calculated
with the equivalent of the Debye formula
\begin{equation}
I(k)=\alpha\sum_{a,b}f_{a}(k)S_{ab}^{(T)}(k)f_{b}(k)\label{I(k)}
\end{equation}
where $f_{a}(k)$ is the form factor for atomic site $a$. Since there
is no such thing as an atom form factor for 2D model atoms, we have
used that of the normalized carbon atom form factor for all sites
(see Fig.S1 in the SI document)
\begin{equation}
f_{a}(k)=f(k)=\frac{f_{C}(k)}{f_{C}(0)}\label{f(k)}
\end{equation}
The coefficient $\alpha$ has a physical meaning in 3D, with $\alpha=r_{0}\rho$,
where $r_{0}=2.8179$ $\cdot10^{-13}$cm is the electronic radius.
If we adopt the same convention in 3D, we can compare the 2D and 3D
spectra, which is what we have done in the sub-Section\ref{subsec:Scattering-properties}.

The various structure factors, intra-molecular correlations and form
factor are illustrated in Fig.S1 of Supplemental Information (SI)
document for the case of ethanol and octanol.

Thermodynamical properties such as the excess internal energy per
particle $E/N$ or the compressibility factor $Z=\beta P/\rho$ are
illustrated for the case of ethanol and octanol in the SI document
in Fig.S2.

\section{Results\label{sec:Results}}

\subsection{The microscopic structure\label{subsec:The-microscopic-structure}}

\begin{figure}[H]
\includegraphics[scale=0.3]{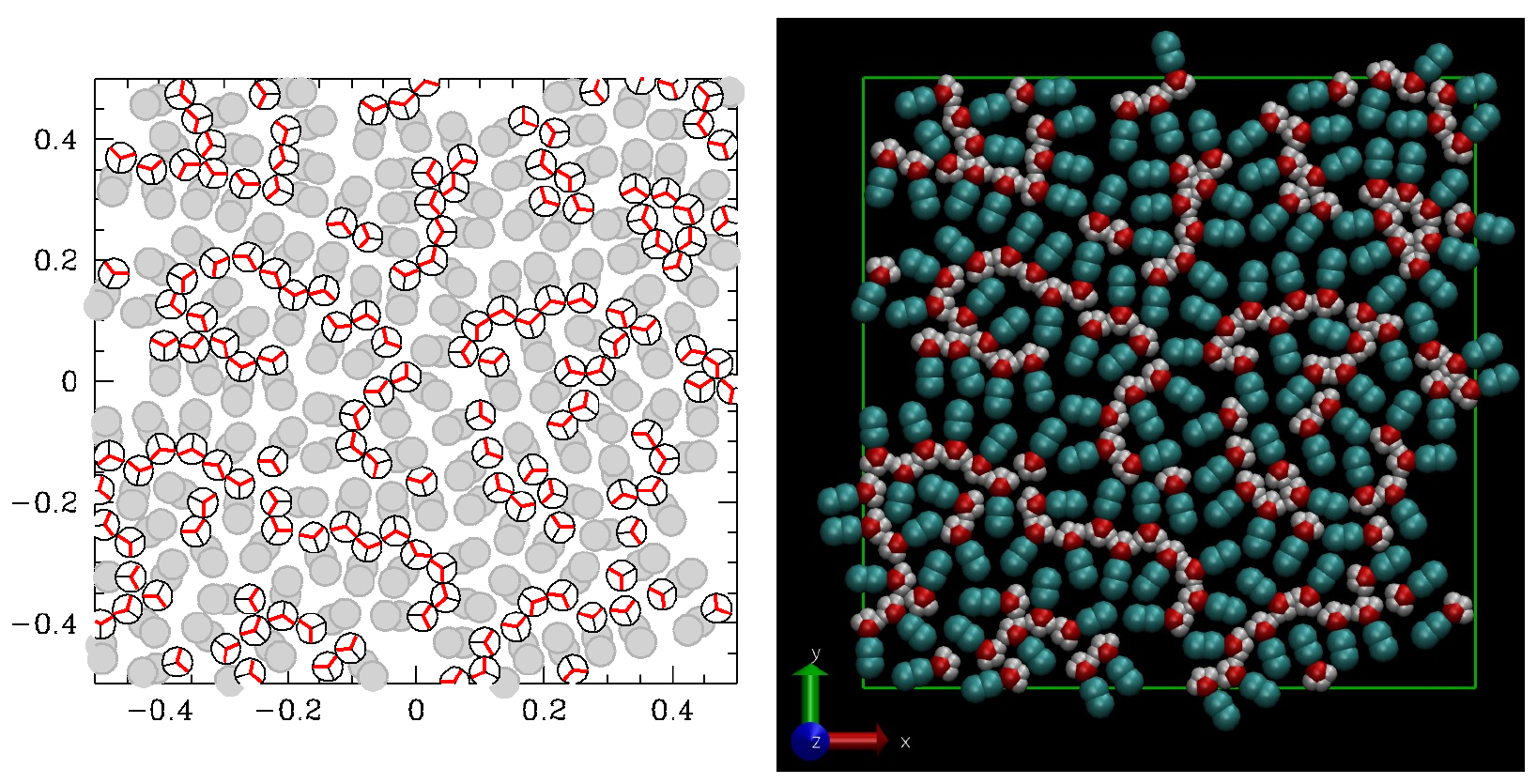}

\caption{Snapshot for 2D ethanol illustrating the hydroxyl head clustering
through MB arm alignments (left panel). This for for a $N=144$ ethanol
molecules system, for $T=2$ and $\rho=0.35$. The carbon atoms are
shown in filled gray discs. The active MB arms are shown in thick
red lines. The right panel is the VMD\cite{VMD} rendering for comparison.
The box border is shown in blue(expand to see).}

\label{FigSnapETH}
\end{figure}

The microscopic structure of the alcohols can be see most directly
by examining snapshots, which illustrates the clustering of the hydroxyl
head groups into short chains of various topologies, mostly chains.
Fig.\ref{FigSnapETH} illustrates for the case of ethanol the MB arm
bonding scheme, which is at the base of the 2D hydrogen bonding model,
and which is well captured by the site-site model. 

While this is for a rather small system of $N=144$ ethanol molecules,
in practice, however, larger systems have been studied, which we examine
now. The most important difference between the two snapshots is that
the left panel shows more clearly that sites seemingly belonging to
a large chain cluster, are not often sufficiently close enough to
be considered as part of that cluster. In other words, although large
chain clusters are apparent, a strict cluster analysis would show
them as broken into smaller entities. This difference is not visible
in the right panel, where only large chain cluster become apparent.
We will consider only this type of snapshot below, since this is closer
to what is seen in the spectral analysis (x-ray scattering for instance),
as will be explained later.

\begin{figure}[H]
\includegraphics[scale=0.3]{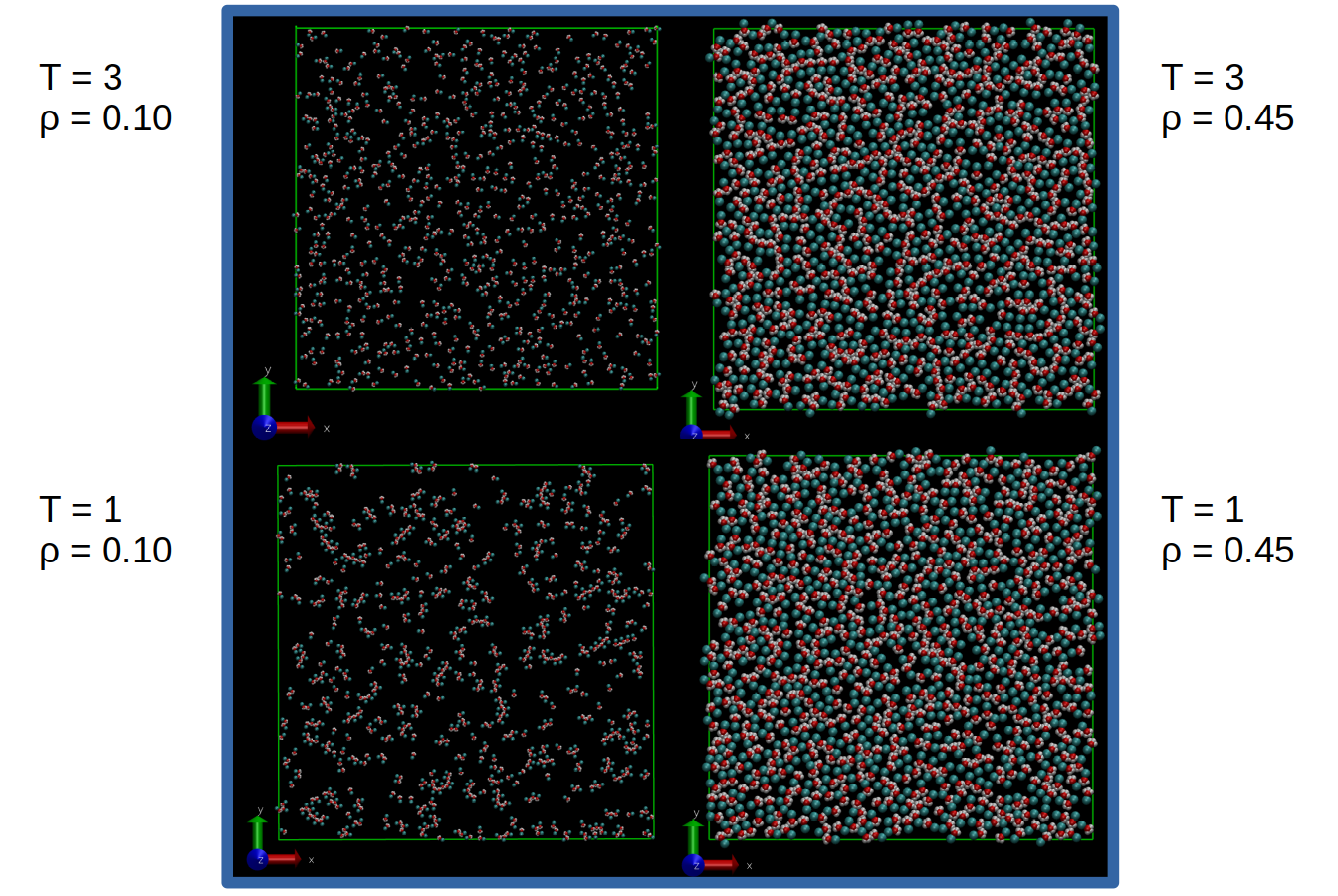}

\caption{Snapshots for methanol for two different temperatures and densities.
The plots are made using VMD \cite{VMD}. Red disc for the hydroxyl
head, with white sites representing the 3 inner bonding sites, and
green sites for the alkyl tail C atoms. These systems concern $N=700$
methanol molecules. The simulation box borders can be seen by expanding
the figure.}

\label{FigSnapMETH}
\end{figure}

\begin{figure}[H]
\includegraphics[scale=0.3]{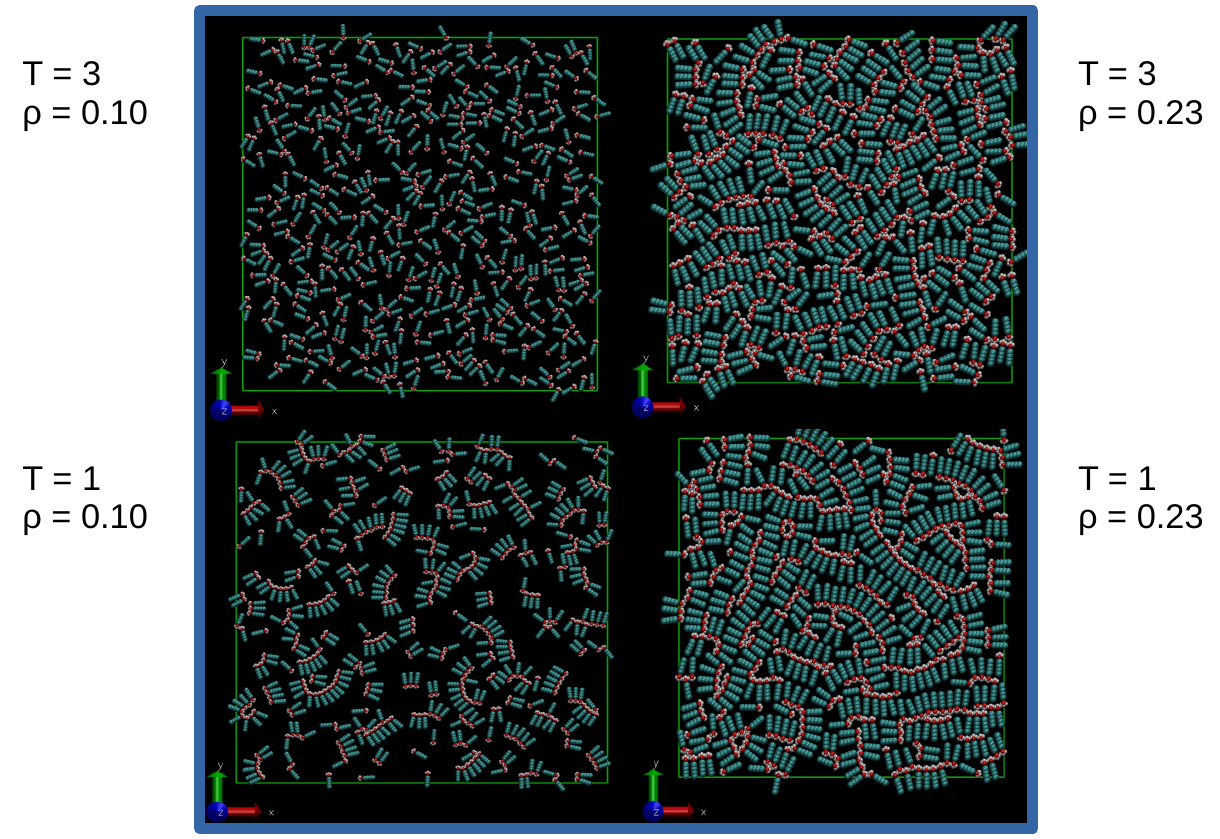}

\caption{Snapshots for butanol for two different temperatures and densities.
The plots details are as in Fig.\ref{FigSnapMETH}. The snapshots
are shown for systems of $N=468$ butanol molecules.}

\label{FigSnapPENT}
\end{figure}

\begin{figure}[H]
\includegraphics[scale=0.3]{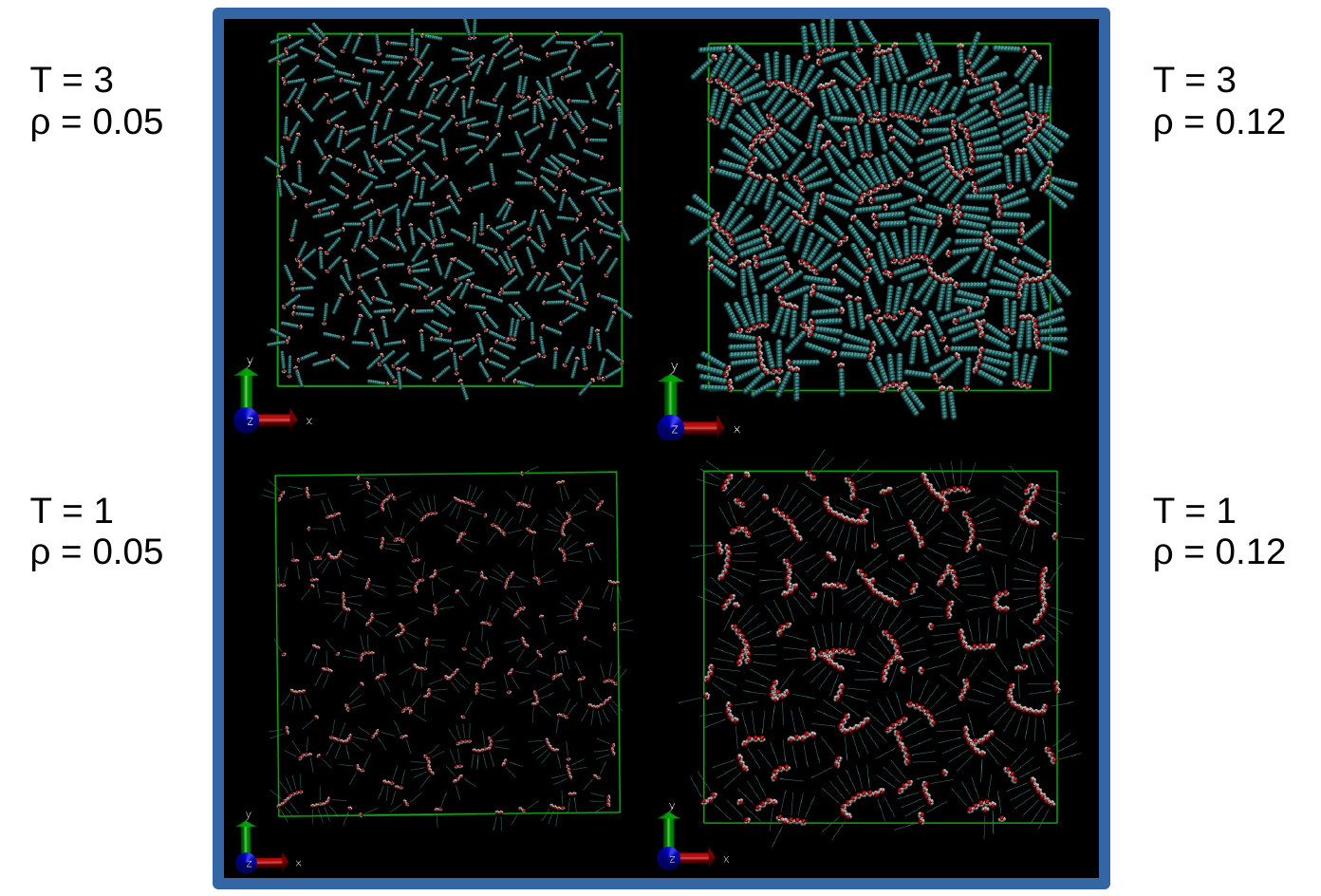}

\caption{Snapshots for octanol for two different temperatures and densities.
The plots details are as in Fig.\ref{FigSnapMETH}. The lower snapshots
show the alkyl tails as sticks, such that the hydroxyl head clusters
are more apparent. The snapshots are shown for systems of $N=336$
octanol molecules.}

\label{FigSnapOCT}
\end{figure}

Figs.\ref{FigSnapMETH}-\ref{FigSnapOCT} show typical configuration
for methanol, pentanol and octanol, thus covering the range from short($n=1$),
medium($n=4$) and long($n=8$) alcohols, where n is the number of
methyl groups. Cases for typical high temperature ($T=3$) and low
temperature ($T=1$) are shown for typical high and low densities.

What is clear, is that for all alcohol models, low density states
look like a gas small clusters, whose size grows with decreasing temperature,
while high density states are more clustered, with a similar temperature
effect. 

Interestingly, this trend is reversed for methanol dense liquid, where
the high temperature state is seen to exhibit larger clusters. This
is due to the smaller alkyl tail, which contributes less to the thermal
induced disorder than when it is larger.

For the case of octanol, only the polar heads are shown, which allows
to see clearly the typical shapes of the clusters, ranging from chains,
to y-shaped branched ones. Circular shapes occur more frequently for
smaller alky tails.

\subsection{Charge ordering through structural properties\label{subsec:Charge-ordering-through}}

The site-site correlation functions $g_{ab}(r)$ for a molecular liquid
are defined as the statistical averaged second moment of the local
density functions $\rho_{a}(\mathbf{r})$ and $\rho_{b}(\mathbf{r})$
for pairs of atoms ($a,b$) as if it was an atomic mixture of all
the atoms in the molecule \cite{HansMac} :
\[
g_{ab}(r)=\frac{1}{\rho_{a}\rho_{b}}<\rho_{a}(\mathbf{r}')\rho_{b}(\mathbf{r}")>
\]
where $r=|\mathbf{r}'-\mathbf{r}"|$ and $\rho_{a}$ is the number
density of atomic species $a$. This definition may be complemented
by the intra-molecular contribution $w_{ab}(r)$ corresponding to
Eq.(\ref{intra}) if one wants to describe the molecular liquid. The
important point is that the site-site function describes fluctuations
of the local densities. In that, they will have particular features
if some atomic clustering features occur. But this clustering is measured
as a form of fluctuation. It is not a measure of clusters themselves.
For instance, the molecule may be considered as a cluster of atoms,
and it is measured by the $w_{ab}(r)$ intra-molecular part. But $g_{ab}(r)$
contains also information as how different atomic contacts occur outside
the molecule.

\begin{figure}[H]
\includegraphics[scale=0.4]{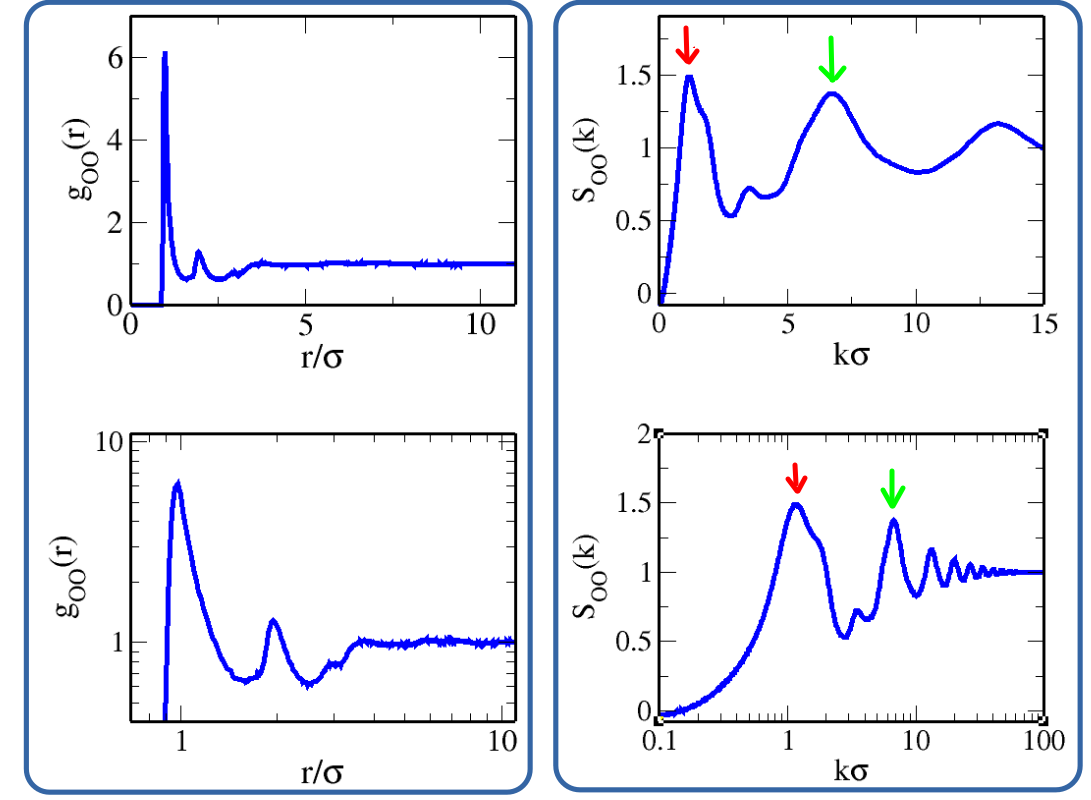}

\caption{Conventions (see text) for structural functions illustrated for the
case of butanol for $T=2$ and $\rho=0.2$. Left panel OO pair correlation
function $g_{OO}(r)$: (top) normal scale, (bottom) log-log scale.
Right panel, structure factor $S_{OO}(k)$: (top) normal scale, (bottom)
semi log scale. The red and green arrows indicate the pre-peak and
main peak positions, respectively. The bottom pictures conventions
will be adopted throughout.}

\label{FigMODE}
\end{figure}

The conventions for displaying correlation functions are illustrated
in Fig.\ref{FigMODE} for the case of butanol for $T=2$ and $\rho=0.2$.
In principle one should show the correlations and structure factors
in normal scale (upper panels). However, because the strong dimerisation
enhanced by the dimensional effect, the main peak of the $g_{ab}(r)$
are very narrow and high, much more than in the 3D case. Similarly,
the pre-peak of the structure factors are as high, if not higher than
the main peak, and the large k oscillatory structure is much more
pronounced than in 3D. For these reasons, it appear convenient to
use log scales for the abscissas (lower panels), which is the convention
that we will adopt throughout.

\subsubsection{Oxygen-Oxygen correlations\label{subsec:Oxygen-Oxygen-correlations}}

\begin{figure}[H]
\includegraphics[scale=0.3]{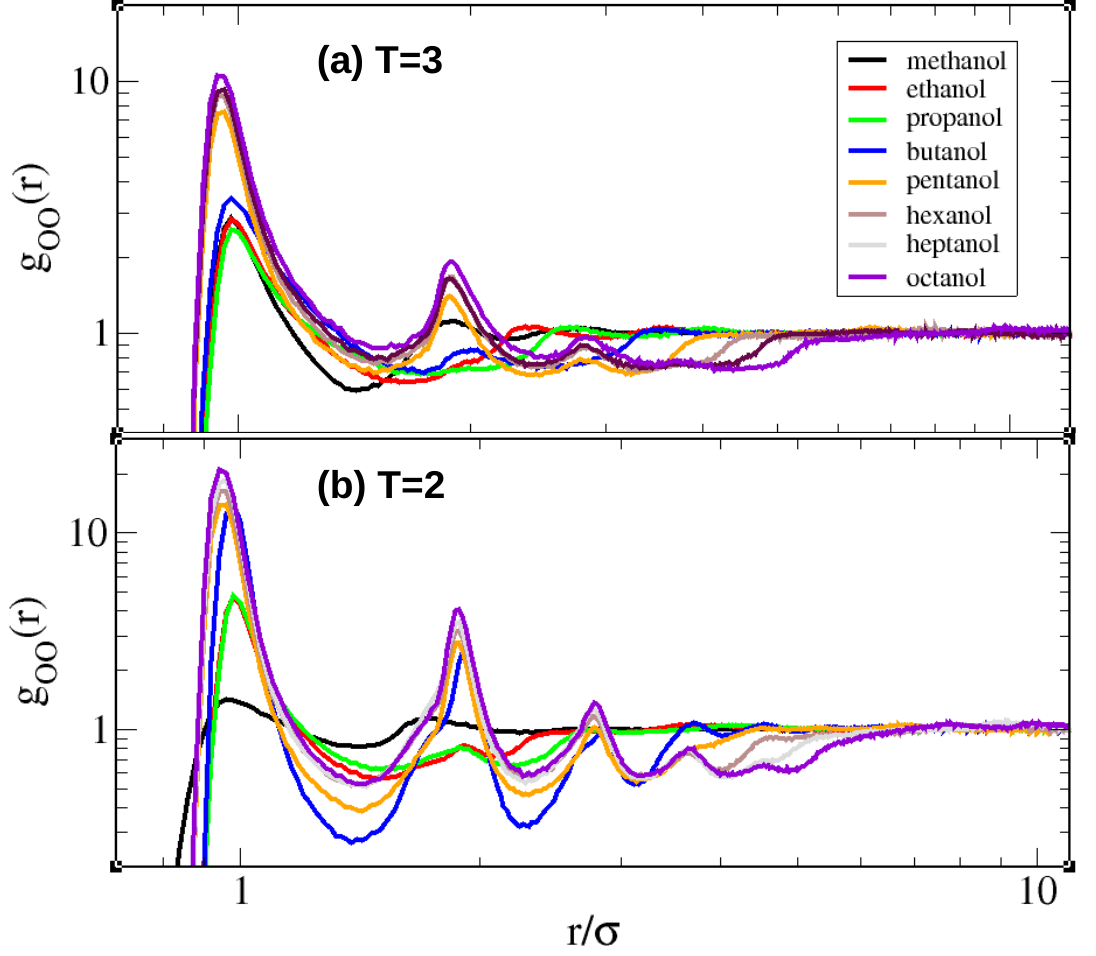}

\caption{Oxygen-oxygen (polar head) correlation functions $g_{OO}(r)$ for
all alcohols for 2 different temperatures $T=3$ (upper panel) and
$T=2$ (lower panel) and for packing fraction $\eta\approx0.465$.
The color conventions are as in the legend in the upper panel.}

\label{FigOOgr}
\end{figure}

Fig.\ref{FigOOgr} shows the polar head (``oxygen'' O atom) correlations
at fixed packing fraction $\eta=0.46475$ ($\eta^{(DR)}=0.69$) for
two different temperatures. The most remarkable feature, aside the
high narrow first peak, is the extended depletion correlation that
follows it, which is more marked with increasing alkyl tails. These
depleted correlations witness the chain-like clustering (which depletes
the O atoms from a spherical distribution around the central O atom).
This is more pronounced that in 3D because of the dimensional restrictions.
In that, the 2D model allows to highlight spectacularly the main driving
effects behind the clustering of alcohols. Lowering the temperature
increases the first peak and extends both the depth and the width
of the depletion correlations (lower panel).

c

\begin{figure}[H]
\includegraphics[scale=0.3]{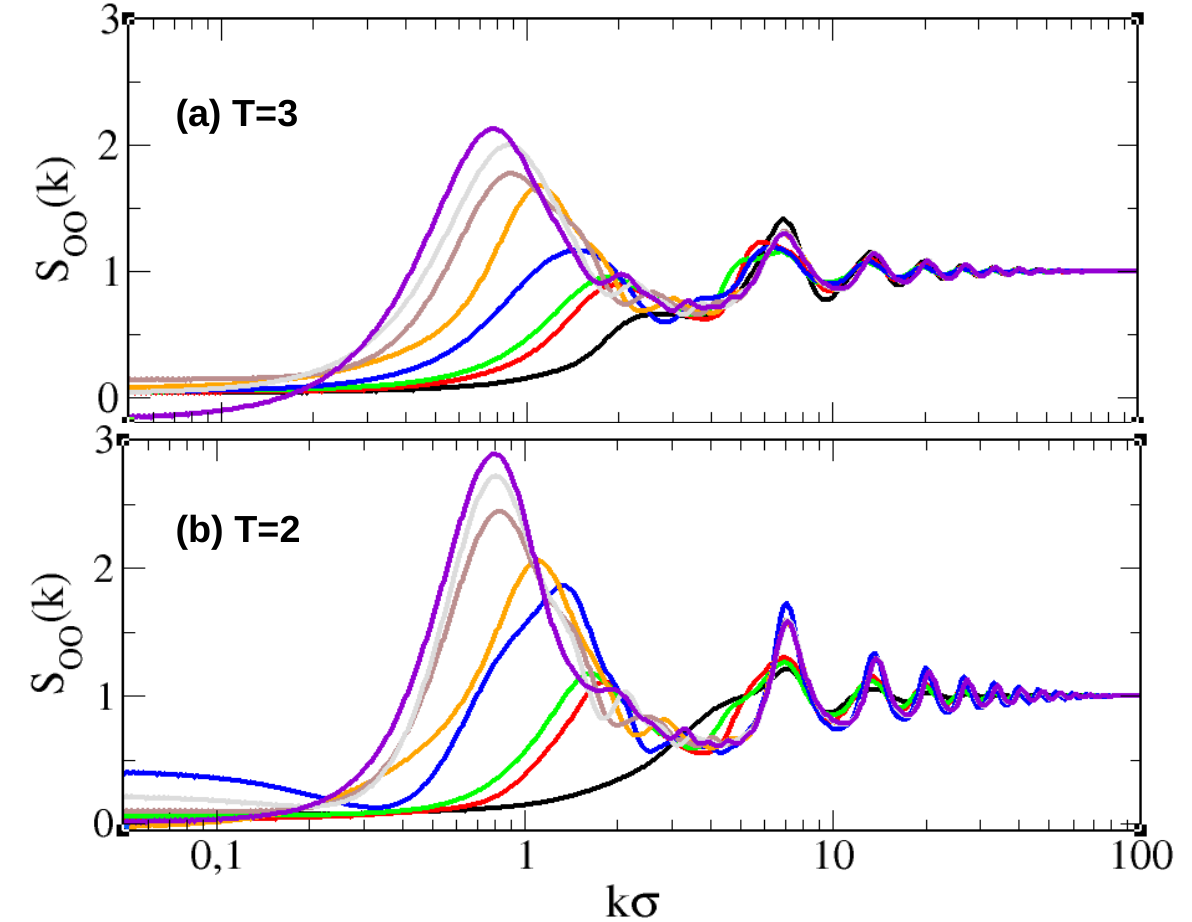}

\caption{Oxygen-oxygen (polar head) structure factors $S_{OO}(k)$ for all
alcohols for 2 different temperatures $T=3$ (upper panel) and $T=2$
(lower panel) and for packing fraction $\eta\approx0.465$. The color
conventions are as in the legend in Fig.\ref{FigOOgr}}

\label{FigOOsk}
\end{figure}

The related structure factors Eq.(\ref{S(k)}) are shown in Fig.\ref{FigOOsk}.
The most relevant feature is the prominent pre-peak, which is even
much higher than the main peak, a feature never encountered in real
alcohols \cite{my_monool}. This is of course a strict consequence
of the depletion correlations in Fig.\ref{FigOOgr}.

The anomalous high temperature induced raise in clustering for methanol,
described in the previous section, can be see here, as the black curve
for T=3 has a more pronounced shoulder peak than for T=2 where the
shoulder moves towards the main peak, hence signaling smaller clustering
trends. Interestingly, the flat shoulder peak for methanol at T=3
is very similar to that in 3D \cite{my_monool}, which shows that
the site model is pertinent enough to represent real 3D alcohols.

\subsubsection{Influence of tail atoms\label{subsec:Influence-of-tail}}

\begin{figure}[H]
\includegraphics[scale=0.3]{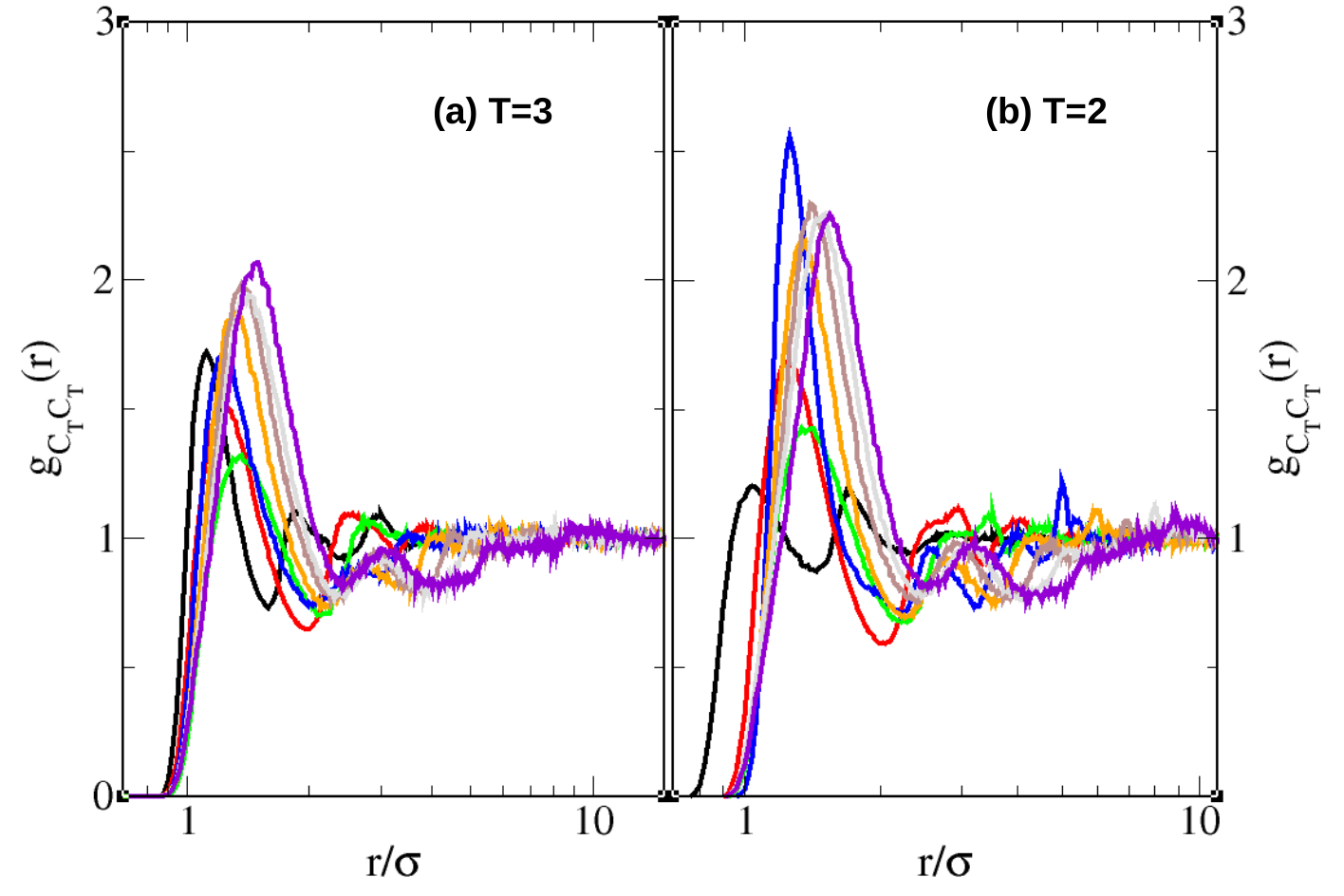}

\caption{Last alkyl tail carbon-carbon pair correlation functions $g_{C_{T}C_{T}}(r)$
for all alcohols for 2 different temperatures $T=3$ (upper panel)
and $T=2$ (lower panel) and for packing fraction $\eta\approx0.465$.
The color conventions are as in the legend in Fig.\ref{FigOOgr}}

\label{FigCCgr}
\end{figure}

Fig.\ref{FigCCgr} shows the correlation functions between the last
carbon atoms for different alcohols. These curves show correlations
that look very much Lennard-Jones like, hence are re presented with
ordinary scale for the abscissa, unlike the previous ones. One can
observe certain amount of depletion, specially for the longer alcohols
(purple curve for instance), which makes sense in view of Figs\ref{FigSnapPENT},\ref{FigSnapOCT},
which show the parallel stacking of the tails, hence a strong chain-like
correlation between carbon atoms of same rank. Another interesting
feature is the correlations for methanol (black curve) which show
a higher first peak and lower second peak for T=3 and both peaks of
similar height for T=2. This is a consequence of the high temperature
clustering inversion discussed above, which is equally reflected in
the carbon correlations.

\begin{figure}[H]
\includegraphics[scale=0.3]{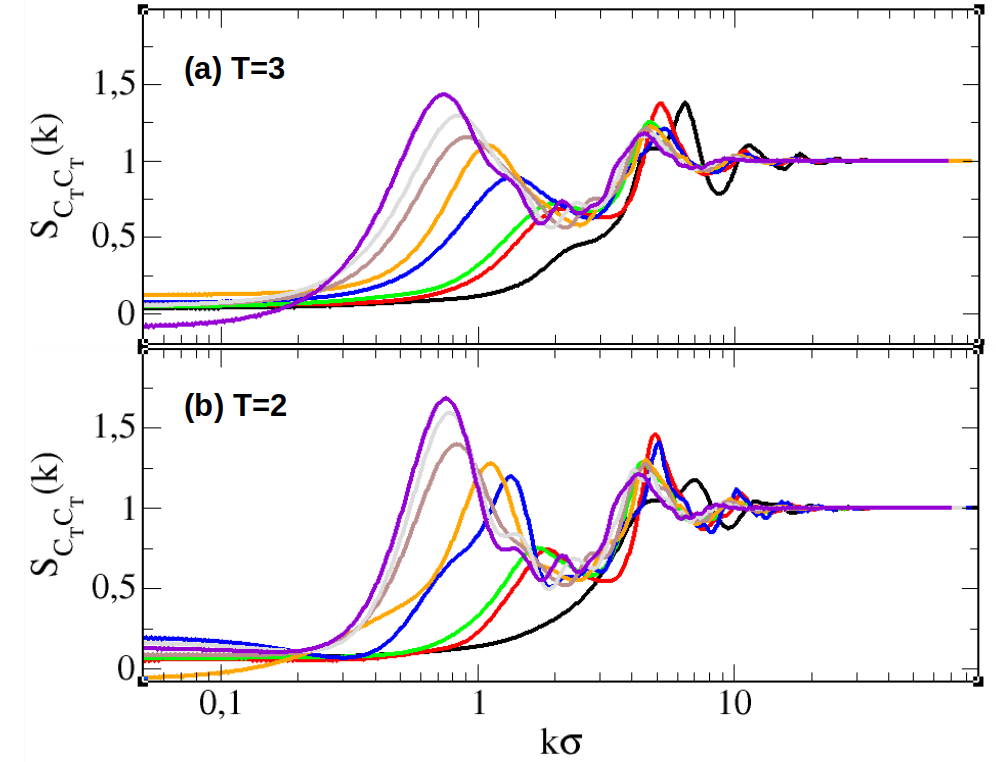}

\caption{Last alkyl tail carbon-carbon structure factors $S_{C_{T}C_{T}}(k)$
for all alcohols for 2 different temperatures $T=3$ (upper panel)
and $T=2$ (lower panel) and for packing fraction $\eta\approx0.465$.
The color conventions are as in the legend in Fig.\ref{FigOOgr}}

\label{FigCCsk}
\end{figure}

Fig.\ref{FigCCsk} shows the corresponding structure factors. Interestingly,
these structure factor show prominent pre-peak, which are also seen
in 3D alcohols, but mostly for longer ones. In the 2D case, all alcohols
show a pre-peak, which is nevertheless not of the same magnitude as
that of the O-O correlations, since these at most equal the main peak
height for most alcohols. Indeed, these correlations are a consequence
of the O-O Hbonding, and the depletion correlations are not as important,
as was observed in Fig.\ref{FigCCgr}. We note that the methanol structure
at T=3 has again this interesting charge order inversion with respect
to T=2.

The depletion correlations, observed for all atom correlations, although
illustrated only in particular cases above, are a direct consequence
of the charge order correlations, where O-P-O chain-like ordering
is observed, similar the hydrogen bond correlations O-H-O in real
alcohols. In the next section we illustrate how such correlations
result from charge ordering.

\subsubsection{Charge-charge correlations\label{subsec:Charge-charge-correlations}}

\begin{figure}[H]
\includegraphics[scale=0.3]{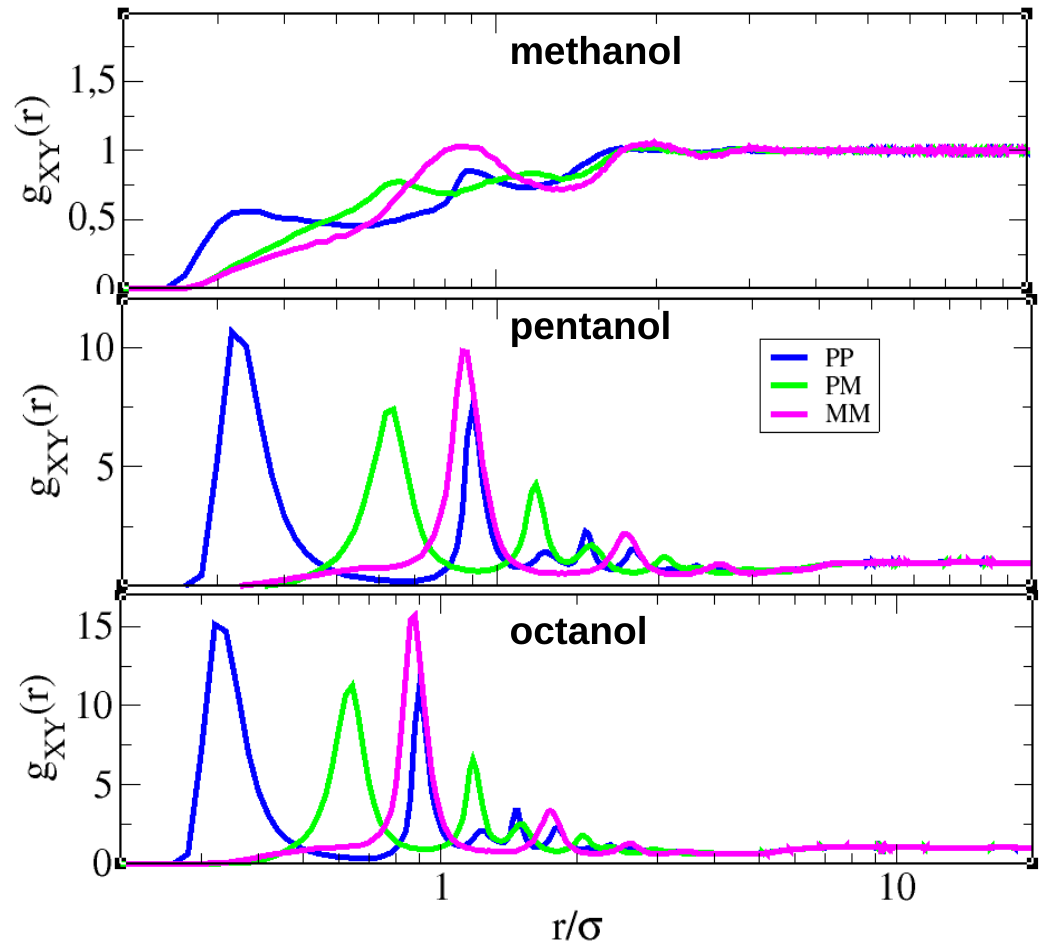}

\caption{Charged sites pair correlations between the blue and magenta sites,
for three typical alcohols from the smallest (upper panel, methanol:
$n=1$) to the longest (lower panel, octanol: $n=8$), with the intermediate
pentanol (middle panel: $n=5$). All curves are for packing fraction
$\eta\approx0.465$ and $T=2$.}

\label{Fig.COgr}
\end{figure}

Fig.\ref{Fig.COgr} shows the pair correlation functions between the
``charged'' sites. The most interesting feature is the fact that
the cross charge correlations are intermediate between the blue and
magenta correlations, as witnessed by the first neighbour peaks. This
cascade results exactly from the charge order, as blue sites tend
to stick together, and the magenta sites are the most repelled. This
is a 2D mimicry of the 3D situation where positive and negative charges
attract each other, contributing to a high peak \cite{myPCCP-CO}
, while the like charge correlations are moved to large distances:
the green curve is out of phase with respect to the blue and magenta
curves which are more or less in-phase.

\begin{figure}[H]
\includegraphics[scale=0.3]{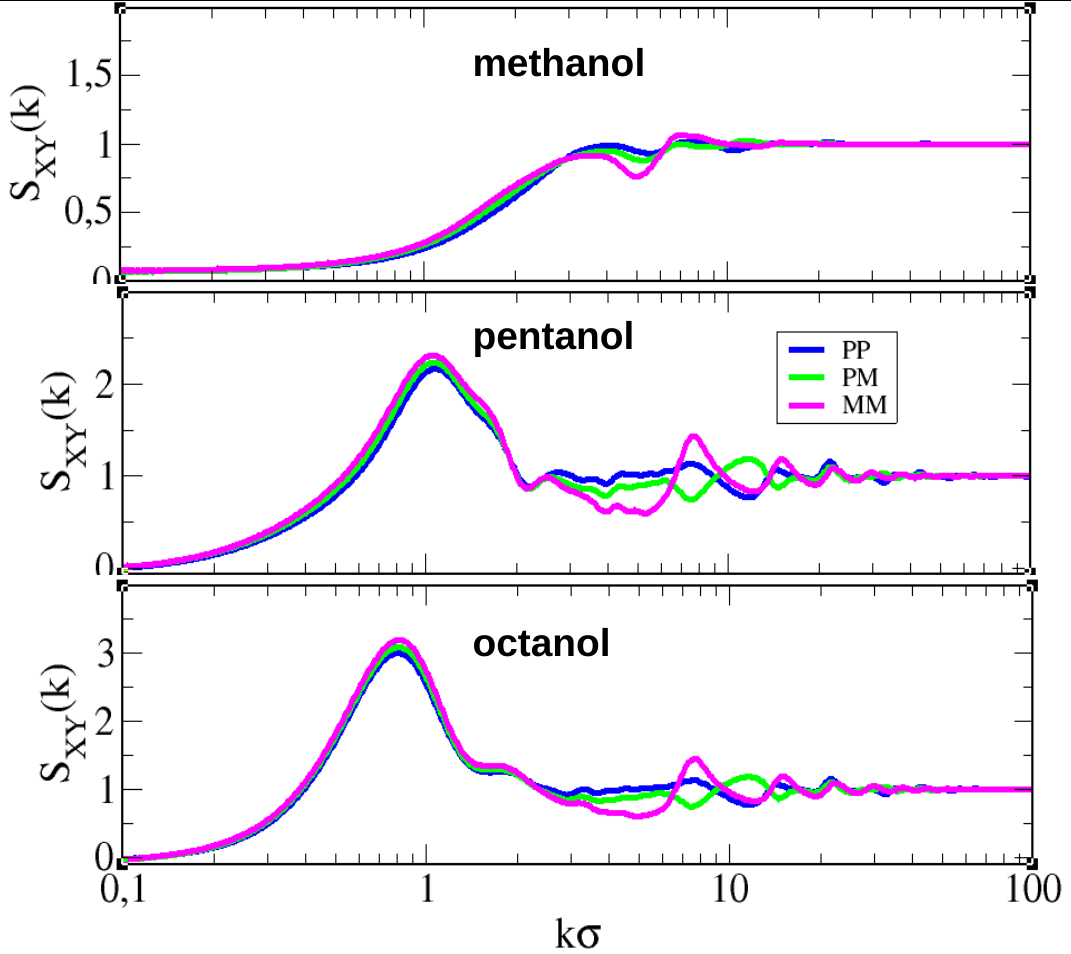}

\caption{Charged sites structure factors between the blue and magenta sites,
for three typical alcohols from the smallest (upper panel, methanol:
$n=1$) to the longest (lower panel, octanol: $n=8$), with the intermediate
pentanol (middle panel: $n=5$), and corresponding to the pair correlation
functions in Fig.\ref{Fig.COgr}.}

\label{FigCOsk}
\end{figure}

Fig.\ref{FigCOsk} shows the resulting structure factors. Interestingly,
although the pair correlations look very dissimilar, their contributions
to the pre-peak are almost identical. The differences are seen in
the main peak positions, which reflect directly those of the corresponding
$g_{XY}(r)$. In particular, in the case of the pentanol and octanol,
it is clearly seen that the cross charge correlations green curve
is an anti-peak to the main peak of the blue and magenta of the like
charge correlations. This is very similar to that observed in the
case of the 3D ionic liquids \cite{myIonicLiq2} , and is the symptom
of the charge ordering process.

We will return to charge ordering when examining the scattering intensities
in Section \ref{subsec:Scattering-properties}.

\subsection{Clustering\label{subsec:Clustering}}

\begin{figure}[H]
\includegraphics[scale=0.4]{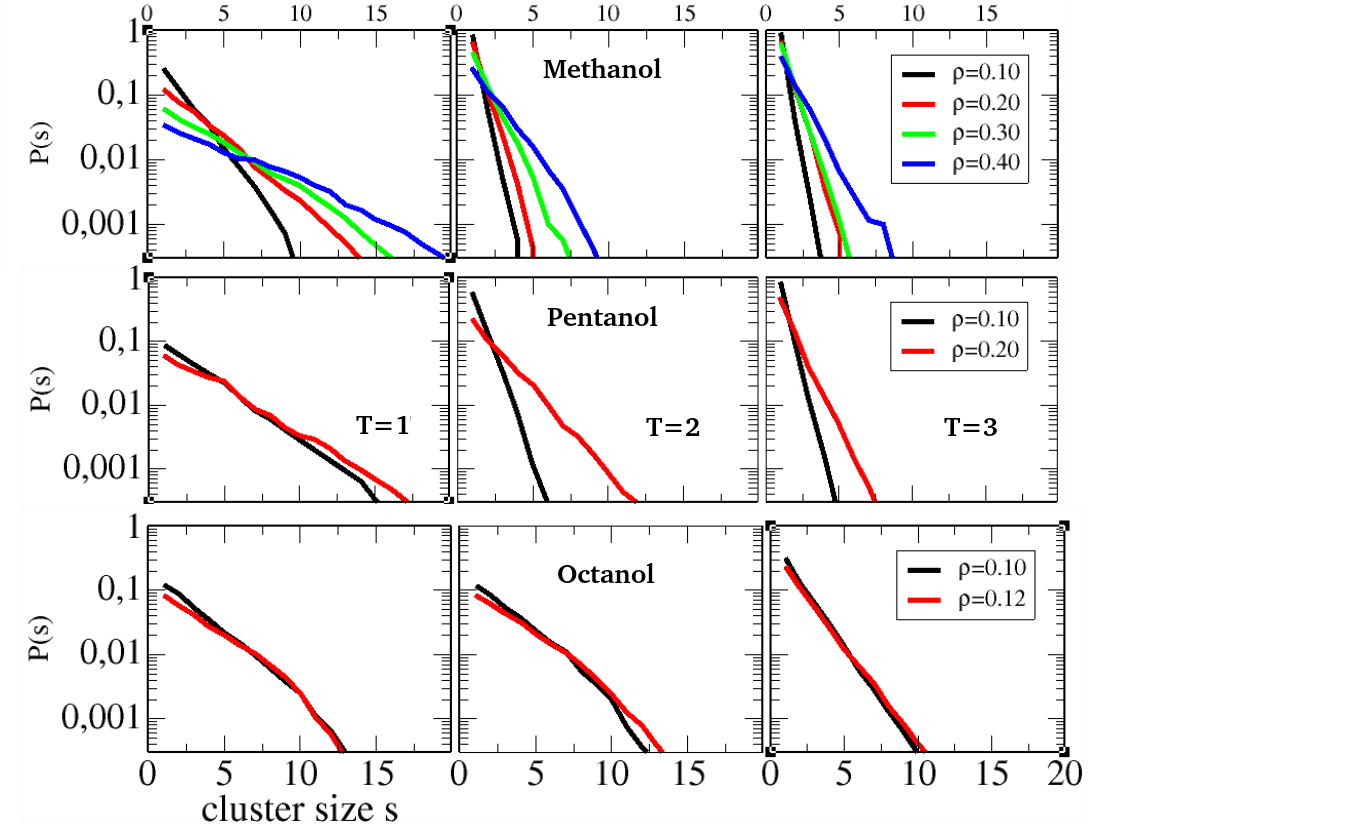}

\caption{Normalized cluster distribution $P(s)$ for methanol (upper 3 panels),
pentanol (middle 3 panels) and octanol (lower 3 panels), for three
different temperatures (a) all left panels T=1, (b) all middle panels
T=2 and (c) all right panels T=3. For each temperature, the evaluated
densities are shown on the inset of the left panels.}

\label{FigClust}
\end{figure}

Instead of the pair correlations which represent cluster fluctuation,
computer simulations allow to probe the clusters themselves, as for
instance the clusters of the $O$-atoms, by building the histogram
of groups of atoms pairs separated by typically the distance which
corresponds to the first minimum of $g_{OO}(r)$ \cite{my_monool}.
The corresponding normalized cluster probabilities $P(s)$ versus
cluster size $s$ are shown in Fig.\ref{FigClust} for 3 different
alcohols, methanol, pentanol and octanol, and for 3 temperatures and
various densities. The first striking feature is that all these curves
is that they very much exponential-like, without however the typical
cluster peak at s$=5-7$ observed in neat alcohols \cite{my_monool}
.

The absence of the cluster peak, when the snapshots show clear charged
group chain aggregate is intriguing. In fact, as mentioned in Section
\ref{subsec:The-microscopic-structure}, a closer look at the chain
clusters show that most of them are broken into smaller aggregates,
and this is what a strict cluster analysis shows. This means that,
even though clear chain-like ``domain'' of the hydroxyl head group
can be observed in simulations, these represent in fact fluctuating
domains containing strict clusters. What the pair correlation functions
measure are these fluctuating ``domains'', just like the visual
analysis would suggest. I am tempted to suggest that it is these labile
domains which represent the true clusters in term of micro-structure.
In real 3D alcohols, the strict clusters are stabilized by dimensional
effects, such as the fact that the alkyl tails surround the polar
head chain clusters and ``block'' their fluctuations, and this produced
the typical cluster peak.

A closer analysis shows trends of $P(s)$ which can be expected, such
as $P(s)$ becomes more exponential at high T because of increasing
disorder, or $P(s)$ for low density decays faster than for high density,
because clusters are larger at high density. These trends are less
marked for larger alcohols, which is equally expected, from the observation
of the snapshots.

\subsection{Scattering properties\label{subsec:Scattering-properties}}

\begin{figure}[H]
\includegraphics[scale=0.3]{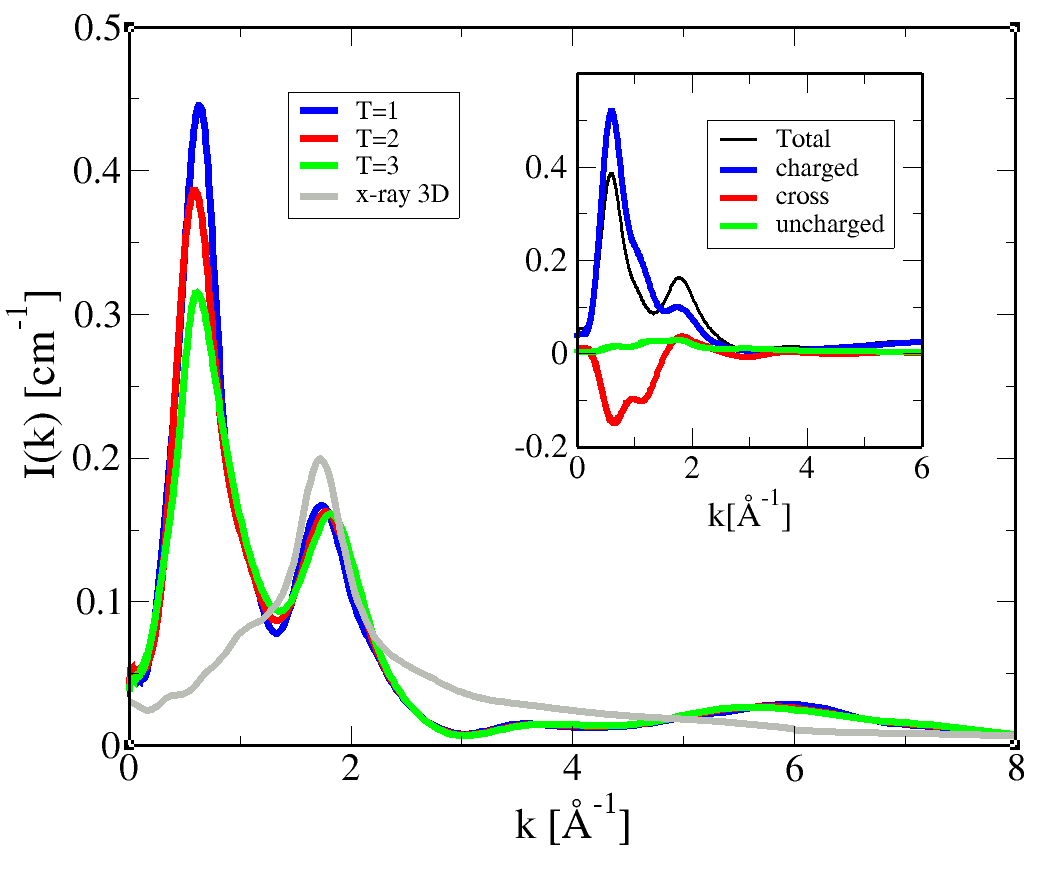}

\caption{Calculated x-ray scattering intensities for the 2D methanol model
for $\rho=0.45$ and 3 different temperatures $T=1,2,3$, and compared
with that from the OPLS model of real ambient condition methanol (gray
curve) from Ref.\cite{my_monool}. The inset shows the partial contributions
to $I(k)$ (black curve) for $T=2$, from charged atomic groups (blue),
uncharged ones (green), as well as the cross contributions (red).}

\label{FigXrayMeth}
\end{figure}

\begin{figure}[H]
\includegraphics[scale=0.3]{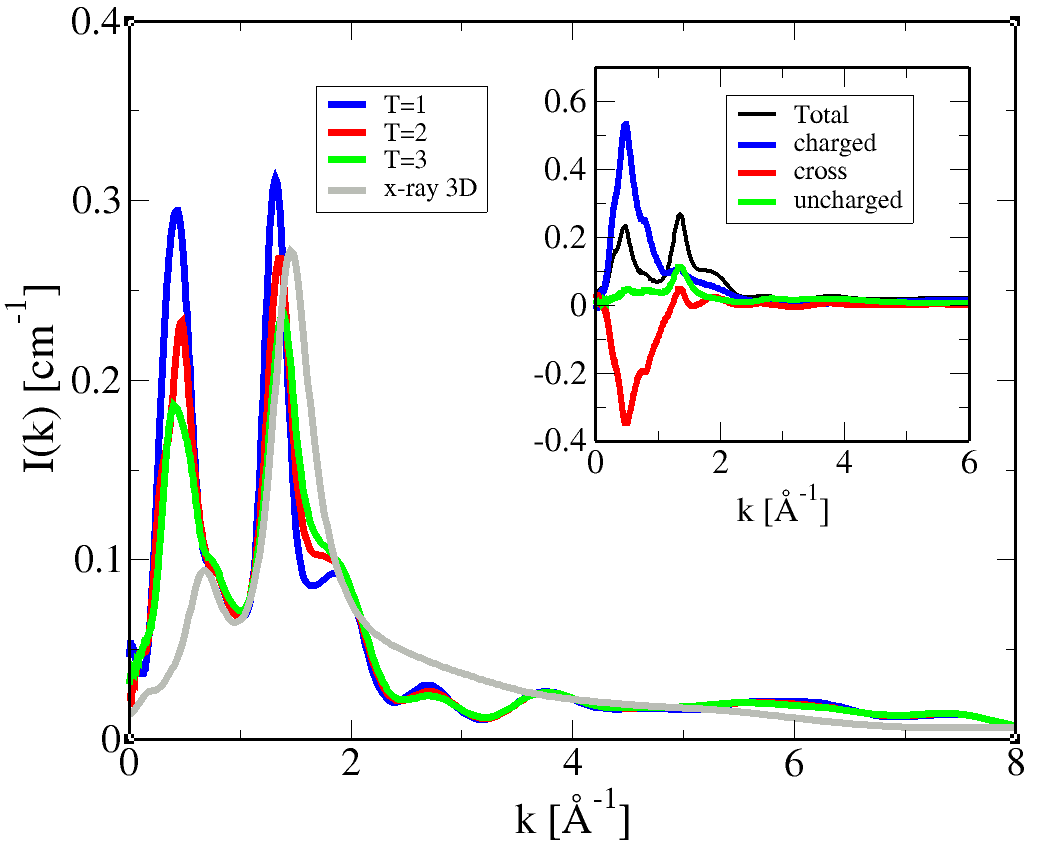}

\caption{Calculated x-ray scattering intensities for the 2D propanol model
for $\rho=0.30$ and 3 different temperatures $T=1,2,3$, and compared
with that from the OPLS model of real ambient condition propanol (gray
curve) from Ref.\cite{my_monool}. The inset shows the partial contributions
to $I(k)$ (black curve) for $T=2$, from charged atomic groups (blue),
uncharged ones (green), as well as the cross contributions (red).}

\label{FigXrayProp}
\end{figure}

Eq.(\ref{I(k)}) allows to calculate the equivalent of x-ray scattering
in real units from the total structure factors calculated for the
model alcohols studied herein. Such x-ray intensities are shown in
Fig.\ref{FigXrayMeth} for methanol and Fig.\ref{FigXrayProp} for
propanol and both are compared with the respective calculated spectra
for the OPLS model of the real alcohols in ambient conditions \cite{my_monool},
shown as gray curve. The $k$ values are arbitrarily scaled with $\sigma=3.5\mathring{A}$,
which is approximately the diameter of the carbon atom in 3D force
field models. This comparison illustrates both the pertinence of the
present 2D models and the differences between real alcohol clustering
and the model, from the perspective of radiation scattering.

The calculated 2D spectra are displayed for 3 different temperatures,
and the amplitudes are seen to increase with decreasing temperatures,
a trend similar to that observed for real alcohols \cite{my_monool}.
The similarities of the main peak contributions at $k\approx1.5\mathring{A}^{-1}$
for both alcohols and their realistic counter part is striking, both
in terms of amplitude and position. In fact, this is not unlike the
comparison of the Lennard-Jones liquid structure factors for 2D and
3D which are equally very much comparable. In other words, it is simply
the expression of packing effect in any dense liquid. The pre-peaks,
however, are a different matter. For both alcohols, the pre-peak positions
are shifted to smaller-$k$ values, indicating that the 2D clusters
are larger than their 3D counter parts. The amplitudes are also much
higher, specially for methanol. If we interpret the amplitude as related
to the density of the cluster, then it make sense since there are
more clustering in 2D than in 3D. This is again particularly true
for methanol, as seen in Fig.\ref{FigSnapMETH}.

c

When the interpretation of the pre-peak position in term of labile
cluster ``size'' is justified, this is not the case for the amplitudes.
Indeed, the insets of both figures show the charged (polar), uncharged
(apolar) and cross contributions to the total spectra 
\begin{equation}
I(k)=I_{\mbox{charged-charged}}(k)+I_{\mbox{uncharged-charged}}(k)+I_{\mbox{cross}}(k)\label{I_partials}
\end{equation}
where each term consists in selecting the appropriate atoms in the
sum in the Debye formula Eq.(\ref{I(k)}), illustrated for for $T=2$.
While it is seen that all 3 contributions are positioned around $k_{P}\approx0.5\mathring{A}^{-1}$,
suggesting a mean cluster size of $R_{c}=2\pi/k_{P}\approx12\mathring{A}$,
the positive and negative contributions to the amplitude do not allow
to decide if the resulting amplitude concerns the density of the clusters.

The separation of $I(k)$ in three different contributions has been
discussed in the case of the 3D liquids \cite{my_monool} . The positive
pre-peak of the charge-charge contribution is essentially due to the
2 factors \cite{myPCCP-CO} , the very high and narrow first neighbour
correlations and the depletion of the second and higher neighbour
(as in Fig.\ref{FigOOgr}). The negative pre-peak, or anti-peak, is
the mostly result of the correlations between the charged and the
uncharged groups being depleted from the first neighbours, for obvious
reasons since these 2 types of atoms do not interact through the pseudo-Coulomb
term in Eq.(\ref{V12}).

The intra-molecular terms contribute importantly to the magnitude
of the pre and anti peaks, as illustrated by Fig.S1 in the SI document.

\section{Discussion and conclusion\label{sec:Discussion-and-conclusion}}

Perhaps the most interesting feature presented in this work is the
fact that a mimicry of the real Coulomb interaction as in the second
term of Eq.(\ref{V12}) is sufficient to produce charge ordering and
correlations effects very similar to that observed in real molecular
liquids and alcohols. Indeed, equation (\ref{V12}) contains two terms
of vastly different magnitudes, specially when the valences $Z_{i}$
of atomic sites $i$ are non-zero. This pseudo-Coulomb model works
very well for modeling 2D ionic liquids \cite{CO1,CO2}, even with
neutral groups attached to charged ones (the so-called room temperature
ionic liquids in the realistic 3D case \cite{margulis}). The model
would allow to mix such ionic species with the present polar species,
and study in two-dimension various interesting phenomena observed
for realistic mixtures. Since the pair interactions between the charged
groups are taken into account, the study of electrolytes and aqueous
mixtures becomes possible, and do not require any extra hypothesis
as in the case of the MB model and variants \cite{tomaz-ionic,MB-ions-jacs}
.

The present work is focused on the micro-structure of 2D associated
liquids, as seen from the site pair correlation functions and corresponding
structure factors. It helps better define the concept of cluster,
as a necessarily fluctuating entity, for which fluctuations play an
essential role in structuring the system. The concept of mean cluster
as referring to tightly bound atomic sites, does not allow to explain
the features observed in the radiation scattering intensity, as was
demonstrated herein. This concept refers to a ``particle'' instead
of referring to a ``fluctuation''. Yet, both representations are
conceivable and the first has been abundantly used in the literature.
I have previously suggested to use the better adapted concept of the
duality between aggregation and fluctuation\cite{myIUPAC}. The present
work serve to illustrate how this concept is embodied in the behaviour
of the density pair correlation functions and their transforms, including
a toy presentation of x-ray scattering in 2D liquids. I expect to
extend this study to model self-assembly, such as that present in
soft matter, where the scattering pre-peak is often used to describe
the self-assembled ``objects'' from strict cluster point of view.
The fluctuation point of view may bring interesting surprises.

\bibliographystyle{jpcb_final.bst}
\bibliography{ssAlc2D}

\end{document}